\documentclass[
 ,twocolumn%
 ,secnumarabic%
,amssymb, amsmath,nobibnotes, aps, prb,showpacs]{revtex4}
\usepackage{bm}%
\usepackage{amsfonts,amsmath,amsopn}
\usepackage{isolatin1}
\usepackage{graphicx}
\usepackage[german]{varioref}
\usepackage[dvips]{epsfig}
\usepackage{graphics}
\usepackage[dvips]{color}
\usepackage{color}
\usepackage{colordvi}
\usepackage{dsfont}
\usepackage{psfrag}
\expandafter\ifx\csname package@font\endcsname\relax\else
 \expandafter\expandafter
 \expandafter\usepackage
 \expandafter\expandafter
 \expandafter{\csname package@font\endcsname}%
\fi

\begin{document}
\newcommand{\be}{\begin{equation}}
\newcommand{\ee}{\end{equation}}
\newcommand{\bea}{\begin{eqnarray}}
\newcommand{\eea}{\end{eqnarray}}
\def\p#1#2{|#1\rangle \langle #2|}
\def\ket#1{|#1\rangle}
\def\bra#1{\langle #1|}
\def\refeq#1{(\ref{#1})}
\def\tb#1{{\overline{{\underline{ #1}}}}}
\def\im{\mbox{Im}}
\def\re{\mbox{Re}}
\def\nn{\nonumber}
\def\t{\mbox{tr}}
\def\sgn{\mbox{sgn}}
\def\Li{\mbox{Li}}
\def\P{\mbox{P}}
\def\d{\mbox d}
\def\i{\int_{-\infty}^{\infty}}
\def\ip{\int_{0}^{\infty}}
\def\mi{\int_{-\infty}^{0}}
\def\A{\mathfrak A}
\def\AA{{\overline{{\mathfrak{A}}}}}
\def\a{\mathfrak a}
\def\aa{{\overline{{\mathfrak{a}}}}}
\def\B{\mathfrak B}
\def\BB{{\overline{{\mathfrak{B}}}}}
\def\b{\mathfrak b}
\def\bb{{\overline{{\mathfrak{b}}}}}
\def\R{\mathcal R}
\def\dm{\mathfrak d}
\def\dd{{\overline{{\mathfrak{d}}}}}
\def\D{\mathfrak D}
\def\DD{{\overline{{\mathfrak{D}}}}}
\def\c{\mathfrak c}
\def\cc{{\overline{{\mathfrak{c}}}}}
\def\C{\mathfrak C}
\def\CC{{\overline{{\mathfrak{C}}}}}
\def\O{\mathcal O}
\def\F{\mathcal F_k}
\def\N{\mathcal N}
\def\I{\mathcal I}
\def\S{\mathcal S}
\def\G{\Gamma}
\def\L{\Lambda}
\def\la{\lambda}
\def\g{\gamma}
\def\al{\alpha}
\def\s{\sigma}
\def\e{\epsilon}
\def\te{\text{e}}
\def\ti{\text{i}}
\def\max{\text{max}}
\def\str{\text{str}}
\def\tr{\text{tr}}
\def\Tr{\text{Tr}}
\def\tC{\text C}
\def\Fo{\mathcal{F}_{1,k}}
\def\Ft{\mathcal{F}_{2,k}}
\def\vs{\varsigma}
\def\l{\left}
\def\r{\right}
\def\up{\uparrow}
\def\down{\downarrow}
\def\u{\underline}
\def\ov{\overline}
\title{$S$=1/2 antiferromagnetic Heisenberg chain with staggered fields:
  Copper pyrimidine and copper benzoate using the density matrix
  renormalization group for transfer matrices}

\author{S. Glocke$^1$, A. Kl\"umper$^1$, H. Rakoto$^2$, J.M. Broto$^2$, A.U.B. Wolter$^3$, S. S\"ullow$^3$}%
\affiliation{$^1$Bergische Universit\"at Wuppertal, Fachbereich Physik, 42097
  Wuppertal, Germany}
\affiliation{$^2$Laboratoire National des Champs Magn\'etiques Puls\'es, 31432 Toulouse, France}
\affiliation{$^3$Institut f\"ur Physik der Kondensierten Materie, TU Braunschweig, 38106 Braunschweig, Germany}
\date{\today}%

\begin{abstract}
We consider the spin-1/2 antiferromagnetic Heisenberg
chain in a staggered magnetic field describing materials such as copper benzoate and copper
pyrimidine dinitrate. Using the density-matrix renormalization group for
transfer matrices (TMRG) we calculate the magnetization of these
materials at finite temperature and arbitrary magnetic field. These results
are in excellent agreement with experimental data, allowing for a
determination of the inhomogeneity parameter c of copper benzoate ($c = 0.043$) and copper
pyrimidine dinitrate ($c= 0.11$). The TMRG approach can be applied to rather
low temperatures yielding singular field and temperature dependences of susceptibilities.
\end{abstract}
\maketitle

One-dimensional quantum magnets have been of theoretical and
experimental interest in recent years because of the rich variety of different
magnetic ground states, such as quantum critical behavior or gaps in the spin
excitation spectra \cite{Hal83,Den97,Sto03}. 
The ideal spin-1/2 antiferromagnetic Heisenberg chain ($S$=1/2 AFHC) with
uniform nearest neighbor exchange coupling is of particular interest, since it
is exactly solvable using the Bethe ansatz (BA) \cite{Bethe31,FGMSK96,KJ00}. Its
ground state is a spin singlet with gapless excitations and quasi-long ranged
ground state correlations, hence even small perturbations can change the
physical properties fundamentally.

In experimental realizations of spin chains like copper
pyrimidine dinitrate
$\left[\mbox{PM}\,\mbox{Cu}(\mbox{NO}_3)_2(\mbox{H}_2\mbox{O})_2\right]_n$
(CuPM) and copper benzoate $\mbox{Cu}(\mbox{C}_6\mbox{H}_5\mbox{COO})_2\cdot3\mbox{H}_2\mbox{O}$
additional terms in the Hamiltonian result from the lack of inversion
symmetry. As a consequence of the residual spin-orbit coupling the
Dzyaloshinskii-Moriya (DM) interaction and an alternating $g$ tensor have to
be taken into account \cite{OA97,AO99}. This gives rise to an effective
staggered field $h_s$ perpendicular to the applied magnetic field $H$. The
Hamiltonian is written as \cite{OA97,AO99}
\be
\label{Hamiltonian}
\hat H= J \sum_i\left[\mathbf{S}_i\mathbf{S}_{i+1} - h_u S_i^z - (-1)^i h_s
  S_i^x\right]
\ee
with the coupling constant $J$, the effective uniform field $h_u = g\mu_B H/J$ 
and the induced effective staggered field $h_s = c\, h_u$. For a given field
axis the effective $g$ and the inhomogeneity parameter $c$ are determined from
the DM interaction $\vec D$ and the alternating $g$ tensor
$\overleftrightarrow{g} = \overleftrightarrow{g_u} \pm
\overleftrightarrow{g_s}$ in the following way \cite{FAGIMMNS00}
$g = \left|\overleftrightarrow{g_u} \vec H\right|/\left|\vec H\right|$ and 
\be
\label{eqn_g_c}
c = \frac{1}{g}\left|\frac{1}{2J} \vec D\times \overleftrightarrow{g_u}\vec H + 
  \overleftrightarrow{g_s}\vec H \right|/\left|\vec H\right|\; .
\ee

In essence, for materials like CuPM and Cu-benzoate, the presence of DM 
interaction and staggered $g$ tensor induces new symmetry breakings with 
respect to the magnetic properties. In terms of principal axes, 
conventionally it is distinguished between the crystallographic unit 
cell with the axes $abc$, the coordinate frame $a^\prime b c^\prime$, where
the uniform part of the $g$ tensor is diagonal, and the system of the principal
magnetic axes 
$a^{\prime\prime} b c^{\prime\prime}$ \cite{AO99,FAGIMMNS00}. For both materials the $g$ tensor and the DM vector have 
been derived previously, as have the different coordinate frames \cite{AO99,FAGIMMNS00}. 
Hence, these quantum spin systems are perfectly suited as model 
compounds for a quantitatively exact comparison between theory and 
experiment. Correspondingly, in recent years various experimental and 
theoretical studies on static or dynamic properties of staggered S=1/2 
AFHC have been carried out \cite{ZWXSY03,Kenzel04,Lou05}, even though data analysis 
was essentially limited to zero temperature. Therefore, in this combined 
theoretical and experimental study we provide a detailed investigation 
of the effect of temperature on thermodynamic properties of staggered 
S=1/2 AFHCs. For the first time, we derive the induced staggered 
magnetization for the staggered S=1/2 AFHC from the low temperature 
regime $k_B T \ll J$ up to the paramagnetic range $k_B T > J$ in fields 
up to saturation.
Since the effective $g$ factor and the $c$ factor depend on the 
orientation of the magnetic field $H$ with respect to the crystal axis 
\refeq{eqn_g_c}, we have to calculate these physical parameters for each 
experimental configuration separately. For this reason, here we 
calculate the magnetization for different crystal alignments by use of 
the transfer matrix renormalization group (TMRG) method and compare these data with experimental data of CuPM
\cite{Wolter03} and Cu-benzoate. The good agreement of numerical and
experimental data allows for the determination of $c$ and $g$ factors. 

\begin{figure}
\includegraphics*[width=7.5cm, height=5.5cm]{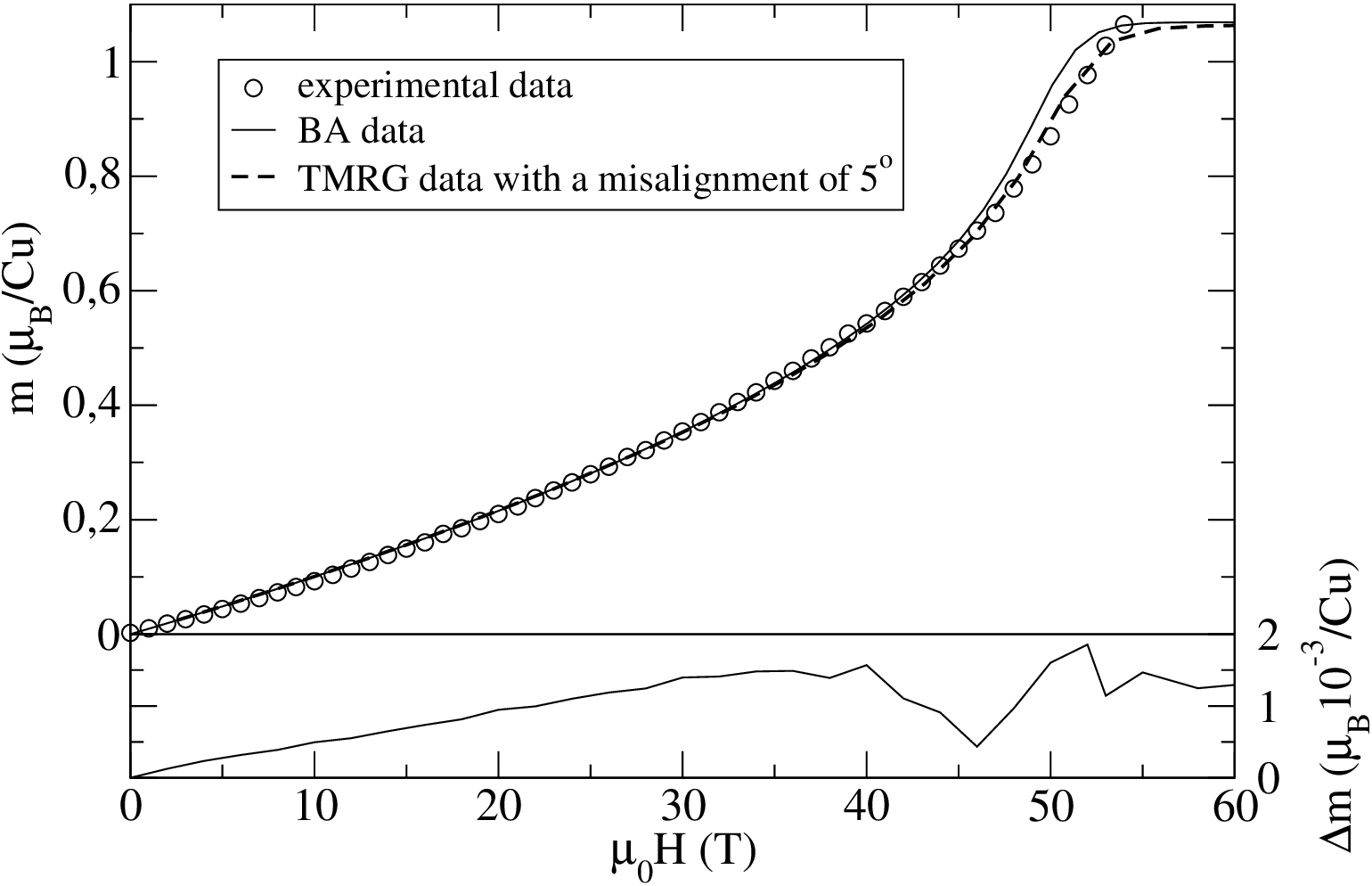}
\caption{Plot of theoretical (solid line) and experimental (open circles)
  magnetization data for CuPM at $T = 1.6$ K along the $a^{\prime\prime}$ direction. Dashed
  line shows TMRG data with a misalignment of $5^\circ$ at $T = 1.6$ K. Note that deviations
  are strongest in the vicinity of the saturation field. Lower plot:
  difference between TMRG and BA data.} 
\label{fig_magn_a16k}
\end{figure}

To study thermodynamical properties at finite temperature the TMRG method
provides a powerful numerical tool, because the thermodynamic limit is
performed exactly and in contrast to quantum Monte Carlo techniques there is
no ``minus sign'' problem. 
The idea of the TMRG method is
to express the partition function $Z$ of a one-dimensional quantum system by
that of an equivalent two-dimensional classical model by a Trotter-Suzuki
mapping \cite{Trotter59,Suzuki85}. For the classical 
model a suitable quantum transfer-matrix (QTM) can be defined which allows for the
calculation of all thermodynamical quantities in the thermodynamic limit by
considering solely the largest eigenvalue of this QTM. Here we 
use the Trotter-Suzuki mapping 
\cite{JK02} yielding
$\label{eqn_part_func}
Z = \lim_{M\to\infty}\Tr\left\{\left[T_1(\epsilon)
  T_2(\epsilon)\right]^{M/2}\right\}\; ,$
where $T_{1,2}(\epsilon) = T_{R,L}\exp\left[-\epsilon H + \O
  (\epsilon^2)\right]$, $\epsilon = \beta/M$ with $\beta$ being the inverse temperature and
$M$ a large integer number. $T_{R,L}$ denote the right- and leftshift
operators, respectively. 
The bulk free energy 
is given by the largest eigenvalue~$\Lambda_0$, 
$f = -T \ln\Lambda_0 \; .$
Expectation values of local operators, like the homogeneous and
the staggered magnetization, can be expressed in terms of the left and right
eigenvectors belonging to the largest eigenvalue.
For achieving lower temperatures the length of the transfer matrix is
increased in imaginary time direction by an application of the infinite density
matrix renormalization group (DMRG) algorithm \cite{DMRG_book}.
In all following calculations we have retained
between 64 and 100 states and used $\epsilon = 0.05$. There are no finite-size
effects as the system size is strictly infinite, however numerical errors due
to the discretization of the imaginary time axis and the DMRG truncation are
generally within the line width used in our plots as has been checked by
comparison with the exact BA data (see Fig. \ref{fig_magn_a16k}). For details
of the algorithm the reader is referred to
Refs. \cite{JK02,DMRG_book,BXG96,Wang_Xiang97,Shibata97}.

In the following we compare our numerical results with high-field 
magnetization data measured at the Laboratoire National des Champs 
Magn\'etiques Puls\'es in Toulouse in pulsed magnetic fields up to 
$\mu_0 H = 53$ T (CuPM) and 38 T (Cu-benzoate), respectively (experimental
details in Ref. \cite{Wolter03}), and with results obtained by means of
$^{13}$C NMR \cite{Wolter05}. 
In the presence of a staggered magnetic field $h_s = c\;h_u$ induced by a
uniform field $h_u$, the magnetization $m$ is given by the superposition of the uniform
$m_u$ and the staggered $m_s$ magnetization components \cite{OA97,AO99}:
\be
\label{eqn_m_phys}
  m = m_u + c\, m_s \; ,
\ee
where $c$ is the above inhomogeneity parameter. The magnetization $m$ can be
measured experimentally. On theory side we can
determine the magnetization $m$ by using \refeq{eqn_m_phys} and calculating
the uniform $m_u$ and the staggered magnetization $m_s$ separately by TMRG
methods.

Recently, CuPM has been identified as a $S$=1/2 AFHC, with a 
magnetic exchange parameter $J/k_B = (36 \pm 0.5)$K \cite{FAGIMMNS00} (in the 
following we use $J/k_B = 36.5$K). In the
uniform $S$=1/2 AFHC model the (magnetic) saturation field is calculated according to the
formula \cite{AO99}
\be
\label{eqn_hc}
H_c = 4 J S / g \mu_B \;.
\ee
The $g$ tensor and the DM vector of CuPM have been derived from ESR
measurements and single-crystal susceptibility studies \cite{FAGIMMNS00}. In
the coordinate frame $a^\prime b c^\prime$ the $g$ tensor takes the form 
\be
\overleftrightarrow{g} = 
         \left(\begin{array}{*{3}{c}}
            2.073&0&0\\
            0&2.149&\pm 0.127\\
            0&\pm 0.127&2.287
            \end{array}
            \right)
	    =\overleftrightarrow{g_u} \pm \overleftrightarrow{g_s} 
\ee
and the DM interaction vector reads 
\be
\vec D = 0.139 J \left(-0.4115,0,0.9114\right)\; .
\ee

If the magnetic field $H$ is directed along the $a^{\prime\prime}$ direction of the CuPM crystal, it will
behave like an ideal $S$=1/2 AFHC ($c = 0$). In this case, from \refeq{eqn_g_c} we obtain
the effective $g$ factor as $g = 2.14$, with saturation field $\mu_0 H_c =
50.7$ T \refeq{eqn_hc}.
In Fig. \ref{fig_magn_a16k} we present the magnetization curve $m$ of CuPM as
function of the magnetic field at T = 1.6 K along the $a^{\prime\prime}$ axis,
as well as the
difference between TMRG and the exact BA results \cite{AK98}. 
Since this difference is always smaller than
$2\cdot 10^{-3}$ the TMRG results describe the exact BA results very well. 
Comparing the experimental data with the TMRG results we find good agreement
with the uniform $S$=1/2 AFHC for magnetic fields up to $\mu_0H < 45$ T. For
higher fields the Dzyaloshinskii-Moriya interaction and the alternating $g$
tensor may take effect on small crystal misalignments, so that CuPM does not
behave like a uniform $S$=1/2 AFHC. Assuming a misalignment of $5^\circ$ of
the magnetic field with respect to the $a^{\prime\prime}$ axis \cite{Wolter03} the physical
parameters have to be changed to $J/k_B = 36.5$~K, $g = 2.13$ and
$c = 0.01$ \refeq{eqn_g_c}.
With these slight modifications the calculated
magnetization curve of CuPM describes the experimental data well, see
Fig. \ref{fig_magn_a16k}.

\begin{figure}
\includegraphics*[width=7.5cm, height=5.5cm]{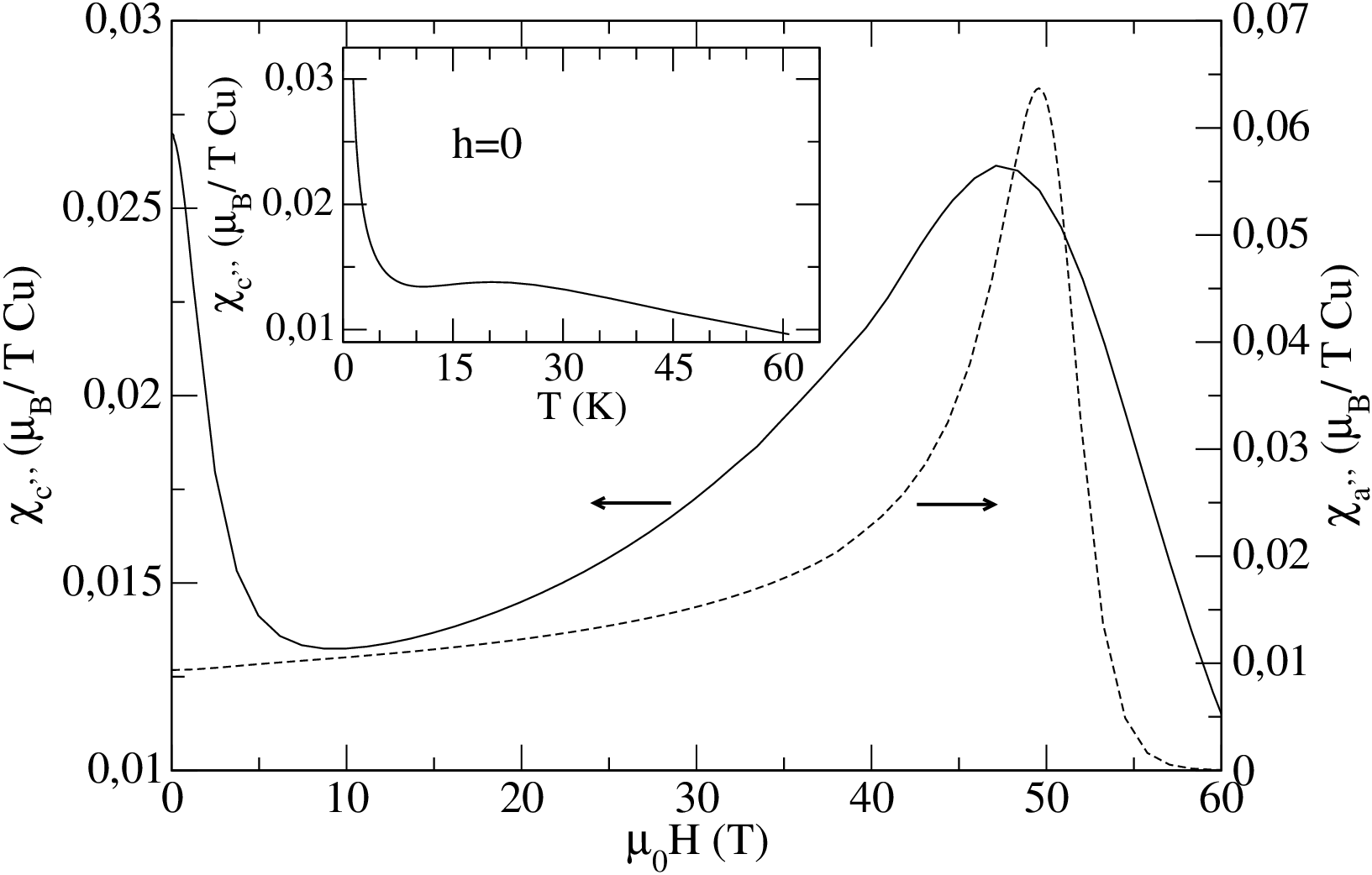}
\caption{Susceptibility $\chi$ of CuPM as a function of $H$ at \mbox{$T = 1.6$ K}
  calculated by TMRG. Solid (dashed) line: susceptibility with magnetic field
  along $c^{\prime\prime}$ ($a^{\prime\prime}$) axis. Inset: $\chi$ of CuPM as a function of $T$ at
  $h = 0$ calculated by TMRG.} 
\label{fig_suszep_CuPM}
\end{figure}
Along the $c^{\prime\prime}$ axis the effect due to the
induced staggered field is largest \cite{FAGIMMNS00}. According to \cite{AO99}
a gap is induced $\Delta \propto h^{2/3}$ with multiplicative logarithmic
corrections. At low $T$ the magnetization receives strong
contributions from the staggered component with singular behavior. For $T=0$ the
dependence $m_s \propto h^{1/3}$ was found and for $h=0$ at low $T$ the
susceptibility $\chi(T) \propto 1/T$ with multiplicative logarithmic
corrections. We stress that this behavior is the result of strong correlations
despite vague similarities with paramagnetic impurities. Susceptibility data
are shown in Fig. \ref{fig_suszep_CuPM}. Note however, the direct application
of DMRG \cite{ZWXSY03} to the Heisenberg chain with full DM-terms does not
give evidences of logarithmic corrections.
\begin{figure}
\includegraphics*[width=7.5cm, height=5.5cm]{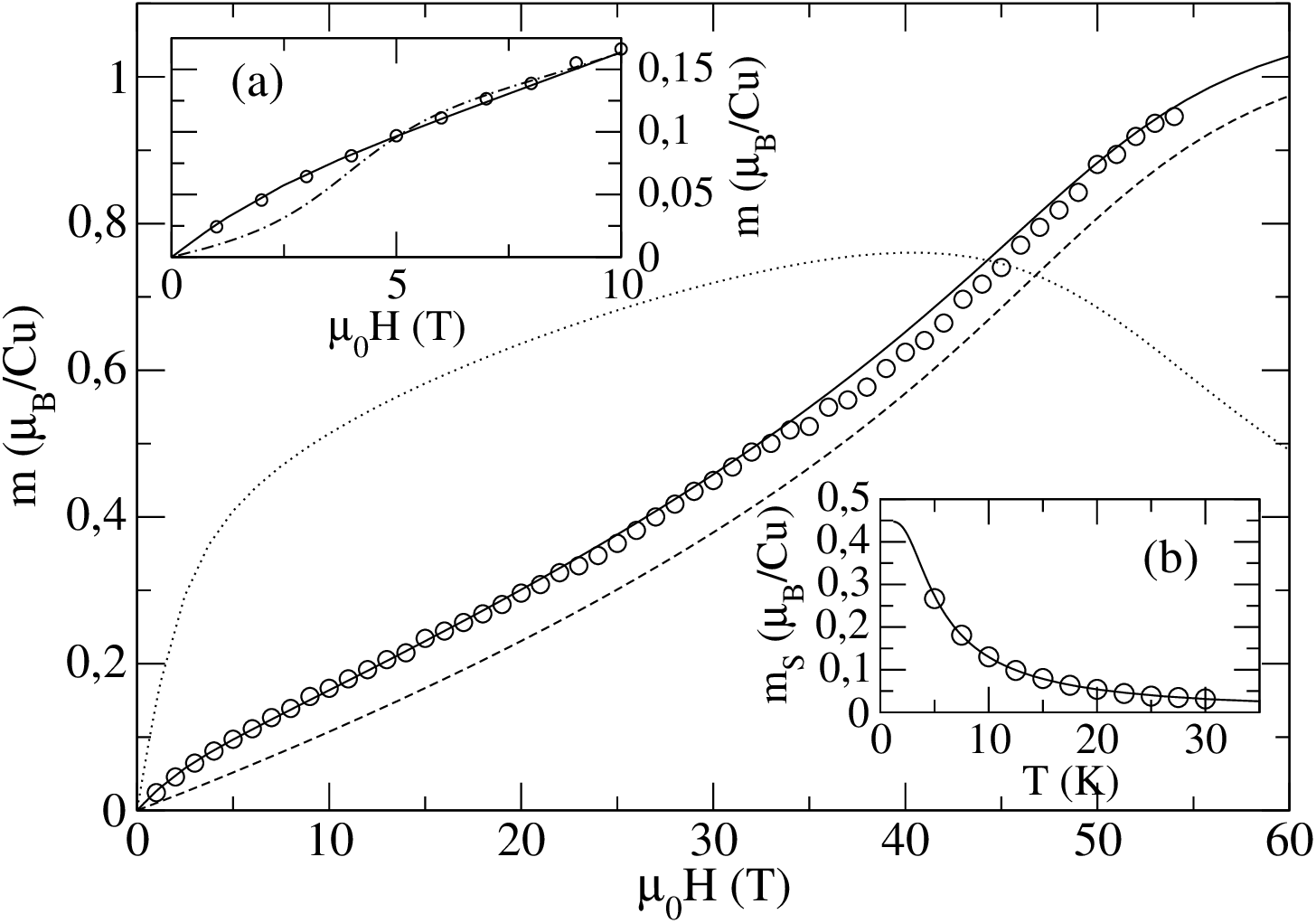}
\caption{Theoretical and experimental magnetization curves for CuPM at $T = 1.6$ K
  along the $c^{\prime\prime}$ direction. Solid line: magnetization $m$ calculated for $c =
  0.11$, dashed line: uniform magnetization $m_u$, dotted line: staggered
  magnetization $m_s$, open circles: experimental data. Inset (a) depicts the
  magnetization for small fields calculated by TMRG and ED
  (dashed-dotted line) in comparison to the experimental data, inset (b)
  displays the temperature dependent staggered magnetization of CuPM for a
  magnetic field $\mu_0H = 9.3$ T applied along the chain axis (solid line: TMRG
  results; open circles: experimental data).} 
\label{fig_magn_011}
\end{figure}
Along this axis we obtain the parameters $c=0.11$ and
$g=2.19$ with saturation field $\mu_0 H_c = 49.6$~T according to
\refeq{eqn_hc}. \mbox{Figure \ref{fig_magn_011}} shows magnetization curves for CuPM with $H$ along the
$c^{\prime\prime}$ axis at temperature $T = 1.6$~K. 
For all magnetic fields the magnetization calculated by TMRG is in good agreement with
the experimental data. In particular at small 
fields, in comparison with the results of exact diagonalization (ED) \cite{Wolter03} the 
TMRG describes the experiment much more accurately, because it has no 
finite-size effects by construction (inset (a) of Fig. \ref{fig_magn_011}). We stress 
that the ED results were calculated for T = 0, 
whereas the TMRG results are calculated at the same finite temperatures 
chosen in the experiment. 

For magnetic field $H$ parallel to the chain axis \cite{FAGIMMNS00} of a CuPM crystal we find
the parameters $c=0.083$ and \mbox{$g = 2.117$}. In inset (b) of
Fig. \ref{fig_magn_011} we present the excellent agreement of the calculated
and the measured \cite{Wolter05} staggered magnetization curves for magnetic
field $\mu_0 H = 9.3$~T. 
Since these results are in good agreement with the experimental data, we
verified the physical parameters for CuPM, especially $c = 0.11$ along the
$c^{\prime\prime}$ axis with accurancy $\pm 0.01$. 
Note that this is consistent with results based 
on ED \cite{Wolter03}, but rules out c=0.08 obtained in [23] from ESR.

Cu-benzoate is another example for a staggered $S$=1/2 AFHC, with a behavior very similar to CuPM. Only, for Cu-benzoate the 
coupling constant is smaller by a factor $\sim 2$ than for CuPM, that is 
$J/k_B = \left(19 \pm 0.5\right)$ K \cite{OA97,AO99}. In consequence, for this
material, in a temperature dependent high magnetic field study, the regimes
from the fully staggered one $k_B T \ll J$ up to the paramagnetic range $k_B T
> J$ in fields up to saturation are accessible and can be compared to the
results from TMRG calculations. From ESR measurements the $g$ tensor in the
$a^{\prime\prime} b c^{\prime\prime}$ coordinate system takes the form:
\be
\overleftrightarrow{g} = 
       \left(\begin{array}{*{3}{c}}
         2.115&\pm0.0190&0.0906\\
         \pm 0.0190&2.059&\pm 0.0495\\
         0.0906&\pm 0.0495&2.316
       \end{array}
       \right)  
\ee
Moreover the DM interaction has been determined \cite{AO99}
\be
 \vec D = J \left(0.13,0,0.02\right)\; .
\ee
Cu-benzoate will behave like a homogeneous $S$=1/2 AFHC ($c=0$) with a
coupling constant $J/k_B = 19$~K, if the magnetic
field is along the $a^{\prime\prime}$ axis. 
\begin{figure}[!ht]
\includegraphics*[width=7.5cm, height=5.5cm]{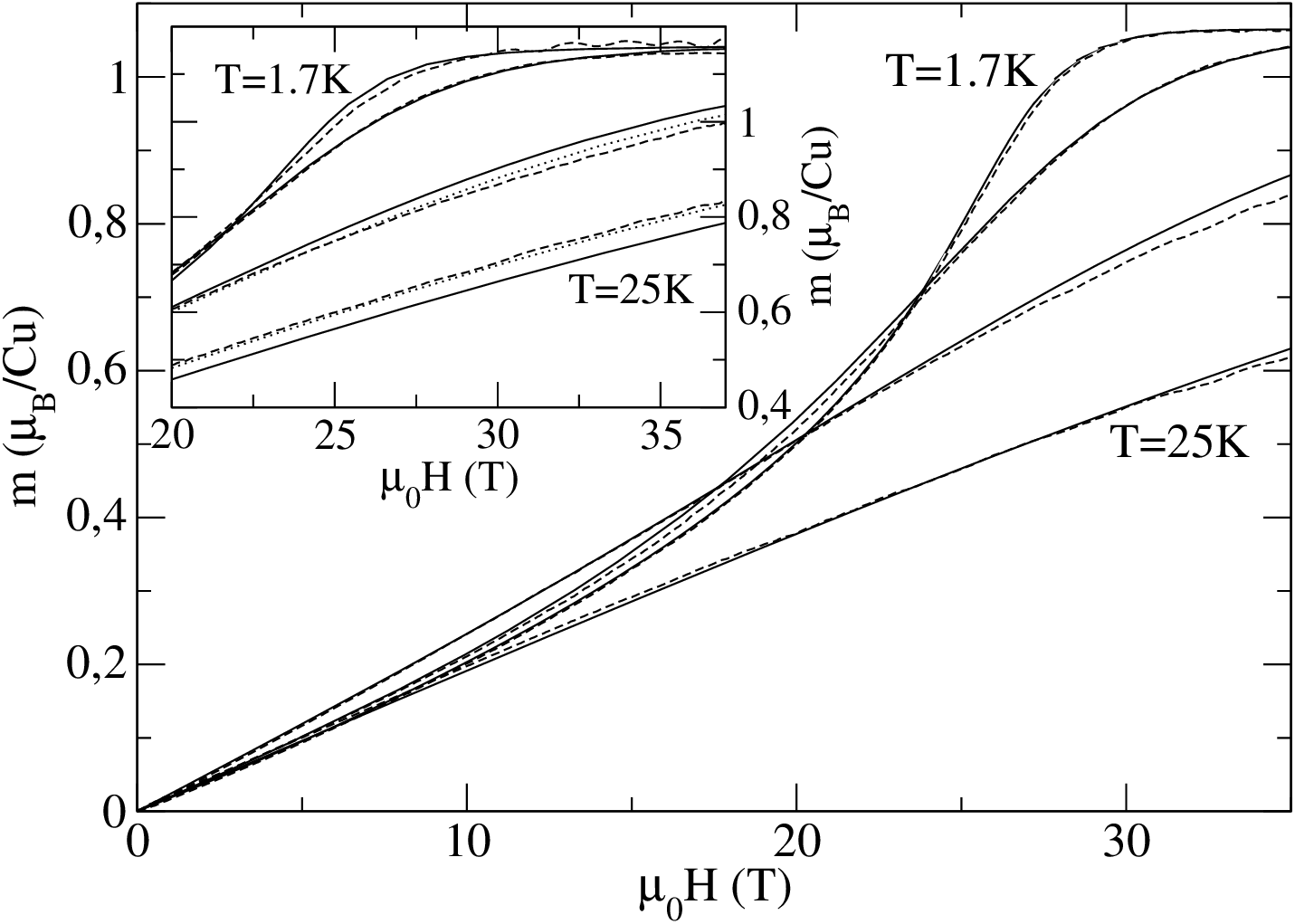} 
\caption{Homogeneous magnetization of Cu-benzoate along $a^{\prime\prime}$ axis at
  different temperatures (1.7K, 4.2K, 12K and 25K). Solid (dashed) lines show
  TMRG (experimental) data. Inset: Magnetization of Cu-benzoate along $c^{\prime\prime}$ axis at different
  temperatures (1.7 K, 4.2 K, 12 K and 25 K). 
  The dotted lines show the
  magnetization at $T = 12.8\mbox{K}$ and $T = 22.5\mbox{K}$ calculated by
  TMRG.}
\label{fig_cuben_a}
\end{figure}
The saturation field is calculated to
$\mu_0 H_c = 26.5$~T by using $g = 2.13$ corresponding to this
axis. In Fig. \ref{fig_cuben_a} we show the magnetization \refeq{eqn_m_phys}
as a function of the magnetic field at different temperatures. Overall, the
TMRG results agree very well with the experimental data obtained in our
high-field magnetization study, especially with respect to the saturation  
magnetization. 
\begin{figure}[b]
\includegraphics*[width=7.5cm, height=5.5cm]{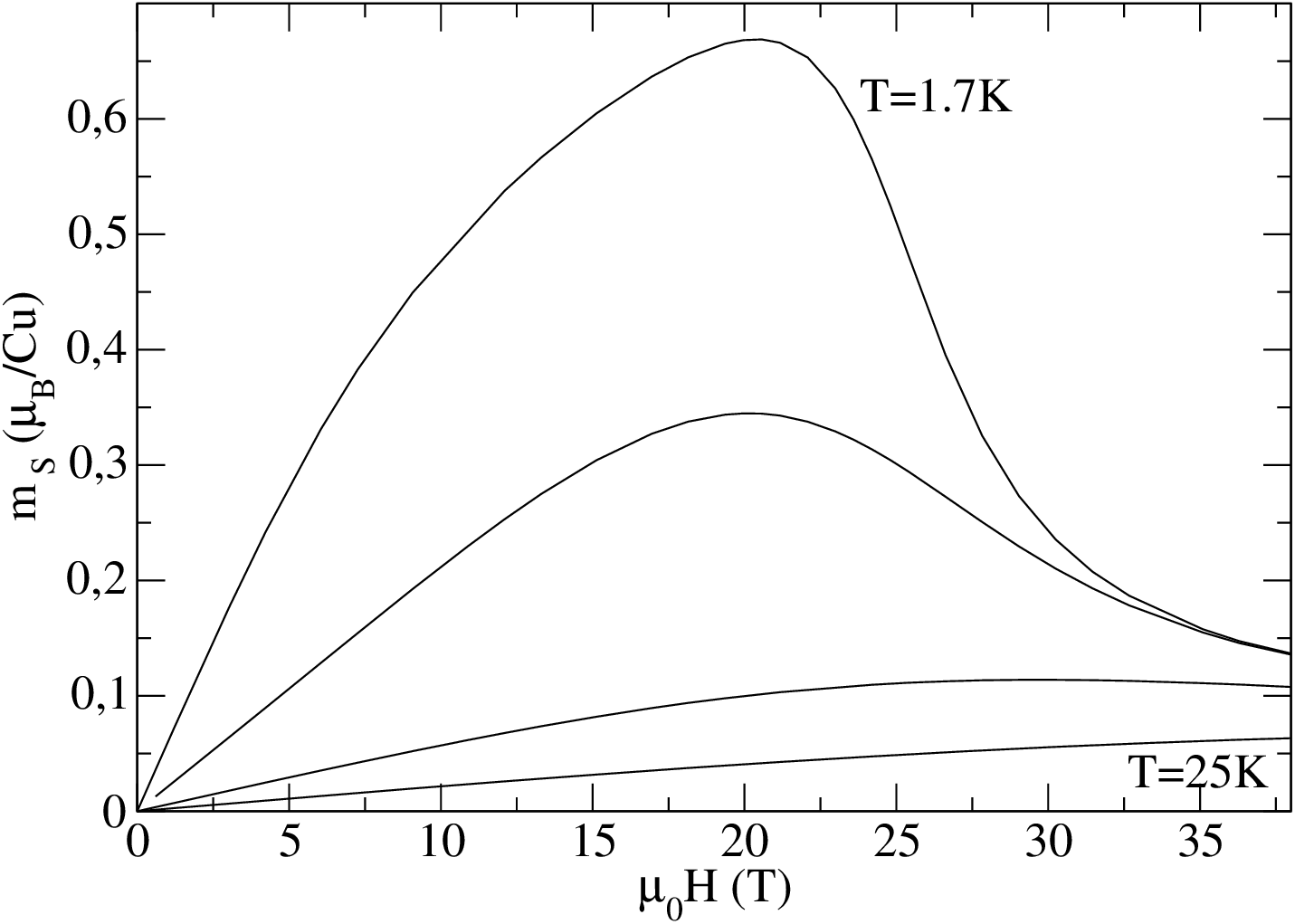} 
\caption{Staggered magnetization of Cu-benzoate along $c^{\prime\prime}$ axis at different
  temperatures (1.7 K, 4.2 K, 12 K and 25 K).} 
\label{fig_cuben_c}
\end{figure}
The field induced gap of Cu-benzoate is largest for magnetic field along
the $c^{\prime\prime}$ axis. In this case we
use $J/k_B = 18.9$~K and calculate the effective $g$ factor to
$g = 2.32$, with a corresponding saturation field
$\mu_0 H_c= 24.2$~T. By using \refeq{eqn_g_c} the $c$ factor is determined to $c
= 0.043$. The inset of Fig. \ref{fig_cuben_a} shows the magnetization
$m$ \refeq{eqn_m_phys}
of Cu-benzoate along the $c^{\prime\prime}$ axis at
different temperatures. Altogether, within experimental 
resolution we find that the magnetization of Cu-benzoate is nicely 
described by TMRG results using $c = 0.043\pm 0.01$ (covering the value
$c=0.034$ resulting from the DM vector used in Ref. \cite{ZWXSY03}). The somewhat larger 
deviations between experiment and theory at higher temperatures ($T = 25$~K) can be ascribed to a larger experimental uncertainty ($\pm 3$~K) in this temperature range.

In Fig. \ref{fig_cuben_c} it can be 
clearly seen that in the low temperature staggered regime ($1.7$~K $= 0.09 
k_B T/J$) there is a prominent staggered component, with a maximum value 
of about 0.6 $\mu_B$/Cu atom at 20 T. This staggered magnetization is 
gradually wiped out as temperature is increased up to 25 K $= 1.3 k_B 
T/J$, {\it viz.}, the conventional paramagnetic regime.

In summary, a comparative analysis of numerical TMRG and 
experimental data for Cu-benzoate and CuPM at low temperatures was 
presented. In our comparison we found very good agreement between our 
numerical and experimental data. The field dependence of the magnetization was
found to show interesting characteristics at high fields corresponding to
saturation and for low field due to quantum criticality of the $S$=1/2 AFHC.

We thank H.-H Klauss for providing the NMR data of inset (b) of
Fig. \ref{fig_magn_011}. This work has been supported by the DFG
under contracts no. SU229/6-1 and KL645/4-2.

\vspace{-0.3cm}
\end{document}